\documentclass[runningheads]{llncs}

\usepackage[utf8]{inputenc}
\usepackage[british]{babel}
\usepackage{graphicx}
\usepackage{amsfonts}
\usepackage{amssymb}
\usepackage{ifsym}

\def\ojoin{\setbox0=\hbox{$\bowtie$}%
  \rule[-.02ex]{.25em}{.4pt}\llap{\rule[\ht0]{.25em}{.4pt}}}
\def\fullouterjoin{\mathbin{\ojoin\mkern-5.8mu\bowtie\mkern-5.8mu\ojoin}}

\begin{document}

\title{\textbf{Understanding the Semantic SQL Transducer\\ (extended version)}}
\titlerunning{Semantic SQL Transducer}
\author{Théo Abgrall \and Enrico Franconi
}
\authorrunning{T. Abgrall, E. Franconi}
\institute{KRDB Research Centre for Knowledge-based Artificial Intelligence\\ Free University of Bozen-Bolzano, Italy\\
\url{https://krdb.eu}\\
\url{theo.abgrall@student.unibz.it}, \url{franconi@inf.unibz.it}
} 

\maketitle

\begin{abstract}
Nowadays we observe an evolving landscape of data management and analytics, emphasising the significance of meticulous data management practices, semantic modelling, and bridging business-technical divides, to optimise data utilisation and enhance value from datasets in modern data environments.
In this paper we introduce and explain the basic formalisation of the Semantic SQL Transducer, a well-founded but practical tool providing the  materialised lossless conceptual view of an arbitrary relational source data, contributing to a knowledge-centric data stack.\\
A talk about this paper is available at \texttt{\url{http://youtu.be/L2uwlsEG8ZE}}

\end{abstract}

\section{Introduction}
\begin{figure}[t]
\centering{\includegraphics[width=\textwidth]{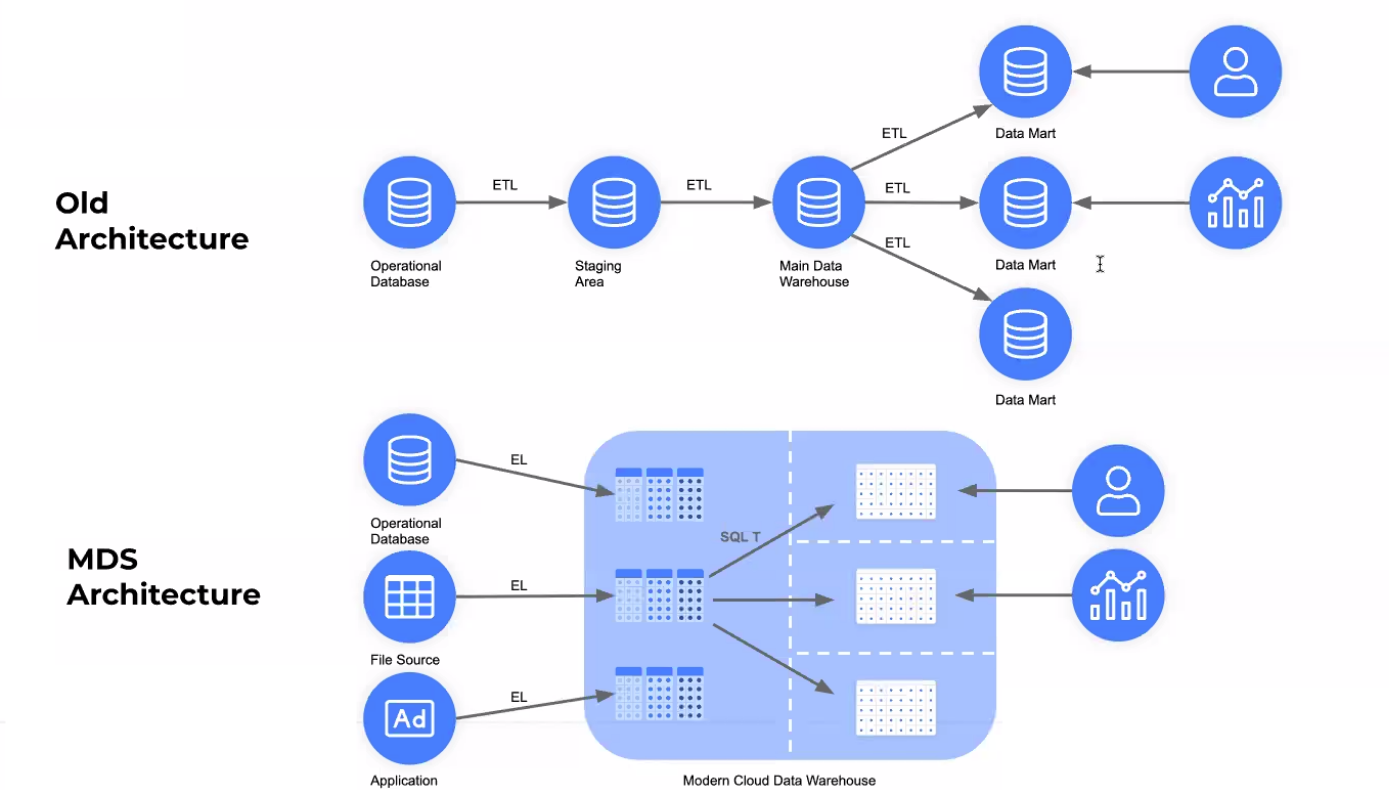}}\\
\centering{\caption{\label{mds}The Modern Data Stack - \textit{source: Fivetran (MDSCON 2020)}}}
\end{figure}

The landscape of data management and analytics is undergoing continuous evolution, aiming to optimise data utilisation, ensure governance, and enhance the value derived from the datasets. Several pivotal concepts shape an enhanced modern data environment, (Figure~\ref{mds}),
 emphasising the significance of robust data preparation, semantic modelling, and bridging the gap between technical and business perspectives.
In~\cite{abgrall:caise-forum-24} we present a thorough analysis of the current trends in data management. Key aspects of these directions include the increasing significance of metadata management for data governance, the necessity of comprehensive semantic enrichment in data contracts and data preparation, the importance of bridging the divide between business problem models and data domains through the integration of semantic mediation, the adoption of a semantic-based declarative transformation process, and the facilitation of seamless data integration and improved interoperability through shared semantic understanding.

\vspace{1ex}
\textbf{Auditability and Data Governance.}
Auditability stands as a critical factor in aiding data analysts to comprehend and model schemas effectively. The emphasis on properly managing metadata supports effective data governance by encompassing the meticulous organisation, quality control, and management of key properties like completeness, consistency, fairness, privacy, provenance, and other data qualities. The focus on enhancing data analyst comprehension and schema modelling is pivotal, highlighting the paramount importance of serious metadata management~\cite{DBLP:conf/itadata/CiacciaMT22}.

\vspace{1ex}
\textbf{Data Preparation.}
Data preparation serves as the foundational pillar upon which successful analytics stands. Tasks encompassed within data prep, such as cleaning, parsing, integrity checks, and data set unification, are crucial. Much like the significance of a solid foundation for monumental structures, a meticulous data preparation phase is indispensable for accurate visualisation, reporting, and forecasting, and it is the step where raw data transforms into actionable insights. Unluckily, the tools that vendors propose for data preparations do not include a serious role of semantics in the pipeline, but only some shallow semantic enrichment such as semantic data types~\cite{DBLP:journals/sigmod/HameedN20}.

\vspace{1ex}
\textbf{Shifting Perspectives on Data Utilisation.}
Historically, the predominant focus in the data domain has been on downstream tasks like visualisation and reporting. However, contemporary trends reveal a substantial shift, where data preparation and meticulous data management consume a significant portion of analysts' time. Modern tools provide a bridge to streamline these preparation-intensive tasks, minimising manual labor and enhancing efficiency~\cite{DBLP:journals/jdiq/ConsoleL23}.

\vspace{1ex}
\textbf{Semantic Modelling and Data-Centric Approaches.}
The traditional approach to data warehousing often resulted in tangled data sets, complicating queries and hindering data trustworthiness. Embracing semantic modelling involves steps like knowledge graphs creation, engineering ownership, data contracts, and subsequent event implementation. Such methodologies ensure that data scientists and engineers spend their time defining and utilising high-quality data, contributing to a more robust and efficient data warehouse. In this context, semantic enhancement plays a critical role in elevating the value and usability of data within industrial contexts. Semantic enhancement isn't merely about structuring data but imbuing it with rich context, relationships, and meaning~\cite{DBLP:journals/semweb/Cudre-Mauroux20}.

\vspace{1ex}
\textbf{Data Catalogs.}
Classical data catalogs and data contracts often fall short in capturing the meaning of the business domain of the companies they are meant to represent. To effectively serve users, catalogs must act as bridges between the user's problem model and the underlying data domain by incorporating semantic understanding. This involves creating ontologies, structured representations of knowledge that enhance the comprehension of data relationships.
When business users attempt to interpret data, they often lack the semantic context necessary for a full understanding. They communicate in their familiar business language, while data appears as a new language tied to the source structures.
Here's where the semantic layer comes into play. This critical component adds a layer of business context to data stored in catalogs, essentially translating it into a language that the business can easily comprehend. The semantic layer also ensures consistency across diverse and heterogeneous data by implementing a common business logic. This conformity empowers businesses to derive meaning from the data coming out of the data stack pipeline~\cite{DBLP:conf/semweb/Dibowski0SHT20}.

\vspace{1ex}
\textbf{Challenges and Solutions in Modern ETL.}
Bill Inmon highlights the evolution of ETL (Extract, Transform, Load) and ELT (Extract, Load, Transform) processes, noting a tendency among vendors and consultants to shift from ETL to ELT. As a consequence, ELT often places the burden of transformation on others, delaying the essential data transformation step. 
``The upshot is – if you wanted to just move some data around, then ELT is your thing. But if you want believable data, then you have to do ETL.
So the choice is yours. Do you want data quickly and easily, that may be essentially unreliable? Or do you want data that forms a firm foundation for – AI, analytics, data mesh, ML, et al?''.
While modern ETL practices have expedited data processing, its procedural nature and lack of a systematic analysis of the data transformation process also brought challenges like incomplete or irrelevant data, tight coupling between services and analytics, and a surge in technical debt. There is the need of a solution involving a return to upfront semantic data definition and design driving a declarative transformation process.

\vspace{1ex}
\textbf{The Importance of Business Literacy.}
Bridging the gap between business and technical perspectives emerges as a crucial necessity. Envisioning and developing semantic warehouses that facilitates enriched communication between business leaders and data practitioners through context-rich data representations serve as the cornerstone for unified and comprehensible data landscapes. As Juan Sequeda says, data modelling without working with subject matter experts/end users is simply a recipe for disaster.

\vspace{1ex}
\textbf{Unifying Ontologies for Data Integration.}
The integration of diverse metadata within enterprises requires a systematic approach. Semantic modelling supports the harmonisation of  disparate data elements within an organisation. A singular, adaptable, and shareable ontology serves as a linchpin, unifying complex data landscapes, fostering cohesive data management strategies, enabling seamless data integration and enhancing interoperability~\cite{DBLP:reference/db/CalvaneseGLLR18}. 

\vspace{1ex}
\textbf{A Pragmatic Approach.}
Dave McComb calls for a pragmatic approach to semantic enhancement, balancing the theoretical underpinnings of ontology development with practical implementations tailored to industrial settings. He advocates for methodologies that facilitate the creation of ontologies that are adaptable, shareable, and capable of evolving alongside dynamic data landscapes.

\vspace{1ex}
\textbf{The Boring Data Stack.}
Joe Reis introduces the concept of the ``Boring Data Stack", signalling a shift in focus from managing underlying technologies to addressing critical yet often neglected aspects like data governance and semantic modelling. He emphasises the  importance of these ``boring" practices, attributing their significance to the era of AI and ML advancements. He points out the  need for organisations to address data quality issues, especially in the context of AI and ML applications, where messy datasets pose significant challenges. Reis envisions a future where data-centric approaches evolve into knowledge-centric ones, emphasising the necessity of robust data governance, management, and semantic modelling practices to achieve this transition effectively. He stresses the criticality of conceptual and logical data modelling in aligning data with the realities of business operations, cautioning against the prevalent tendency to solely focus on physical data models disconnected from business needs.

\vspace{1ex}
Given this context, we believe a contribution to support proper semantic modelling within a data preparation pipeline is badly needed. In this paper we introduce and explain the basic formalisation of the \emph{Semantic SQL Transducer}, a well-founded but practical tool providing the  materialised \emph{lossless} and possibly conceptual view of an arbitrary relational  data, contributing to a knowledge-centric data stack. The Semantic SQL Transducer can be seen as a seamless semantic wrapper around arbitrary relational data at any stage of the data stack, independently on its architecture. The advantage of this technology is that it can be seen as a replacement of the data it models, providing a restructuring of the data according to its restructured (and possibly conceptual) model as a standard SQL database, which can be therefore queried, updated, transformed. It can be used also to replace procedural data transformation tasks with semantic-based declarative executable specification of the transformation task, guaranteeing the losslessness of the transformation itself. By restructuring the relational data directly in its conceptual model, the transducer provides a conceptual access to the data, providing to business users and data analysts the right understanding of the available data.

\section{The role of a Semantic SQL Transducer}
\label{sec:role}

Our contribution to a knowledge-centric data stack consists in a SQL-based tool (supported by a design methodology) providing the  materialised \emph{lossless} conceptual view of an arbitrary relational source data. Such materialised conceptual view can be queried and updated using the knowledge vocabulary, with virtually no overhead with respect to the original source data. Updates to the materialised conceptual view are replicated instantaneously to the source data (e.g., to push a semantic-conscious data cleaning update to the source), and updates to the source data are replicated instantaneously to the materialised conceptual view providing always a fresh view of the source. The materialised view is the conceptual lossless mirror of the source data, and it acts as a mediator by providing the conceptual API for a complete access and change to the source data. 

\begin{figure}[t]
\centering{\includegraphics[width=\textwidth]{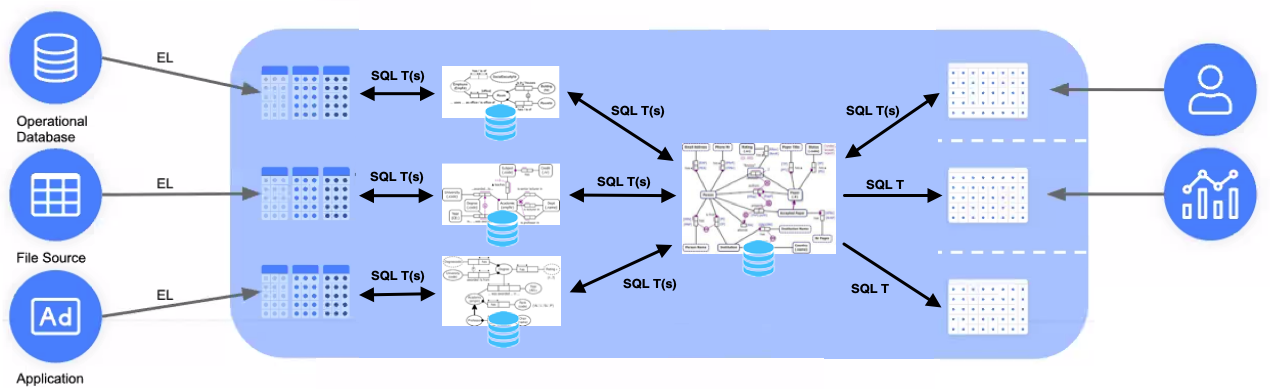}}\\
\centering{\caption{\label{smds}A Semantic Data Stack}}
\end{figure}

Our \emph{Semantic SQL Transducer}
is based completely on standard SQL technology, it can be deployed on any SQL platform, and it does not require any additional tool or code to work.
%
%
The transducer can exactly translate legacy SQL queries and updates over the source schema to SQL queries and updates over the conceptual schema, and it can exactly translate analytical SQL queries and updates over the conceptual schema to SQL queries and updates over the source schema. By using semantic SQL transducers within a data prep process as semantic wrappers around the data sources (e.g., see Figure~\ref{smds}), the ``transform'' part of the ETL process can operate over semantically well defined entities and relationships.
The transducer supports transactions, to systematically guarantee semantic integrity and consistency of both the source and its conceptual model.

In order to represent the exact semantics of the source data, the Semantic SQL Transducer supports several popular conceptual data models: ERD, ORM, UML Class Diagrams, Property Graphs Schemas, Knowledge Graphs. It is based on several years of theoretical research by us on the formalisation of the connection between conceptual data models and relational databases, and on the formalisation of core SQL~\cite{DBLP:conf/dmdw/FranconiS99,lubyte2009automatic,DBLP:books/sp/18/CalvaneseF18,DBLP:conf/adbis/Abgrall22,abgrall:24}. A rigorous methodology to properly design a semantic SQL transducer given the data sources has been studied and experimented~\cite{DBLP:conf/aike/NdefoF19,ndefo:franconi:ijsc-20}; we are developing several tools to support it, which are not yet publicly available.

We believe that our pragmatic approach to semantic enrichment provides a useful knowledge-based core element which can be embedded within many different data architectures, improving on the issues emphasised above in the direction of making the boring data stack more exciting: auditing and data governance now are based on a clear semantic view of the data, supporting a more transparent environment to exploit business literacy; during data preparation, semantic integrity checks can be defined over the conceptual structures, and data cleaning can be enforced at the level of the conceptual view; data integration and entity recognition are now semantic-based, and the presence of a unified conceptual model exactly capturing the diverse data sources supports the harmonisation of the disparate data elements within the enterprise; the gap between data management and data analysis is reduced, being mediated by the transducers, reducing therefore the risk of a technical debt; data scientists operate over high-quality data, contributing to a more robust and efficient data warehouse.

Note that there are many semantic-based approaches that introduce an intermediate layer between the data layer and the business layer, which should be compared the proposal, such as enterprise data fabrics, data meshes, data lakes, etc. It is important to observe that the proposed Semantic SQL Transducer is orthogonal to any of such architectural choices. The suggested semantic data stack in Figure~\ref{smds} above had the only purpose how the Semantic SQL Transducer could support semantic enrichment in a modern data stack. The unicity of our proposal lies in the fact that it can losslessly "present" the data in a restructured way, possibly according to its conceptual schema, always using SQL as the foundational formalism.


\section{Inside the Semantic SQL Transducer}
\label{sec:inside}

\begin{figure}[t]
\begin{minipage}{.74\textwidth}
\includegraphics[width=\textwidth]{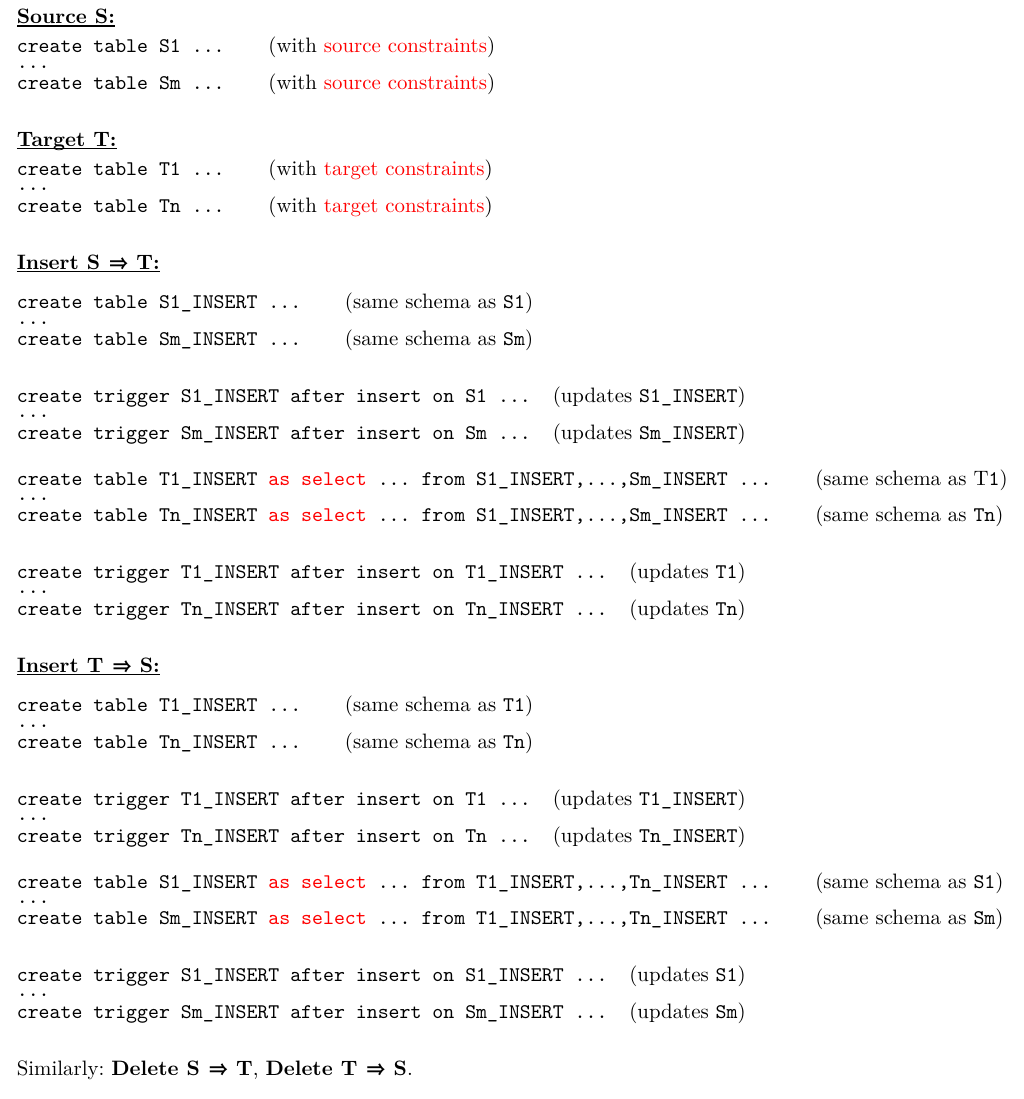}
\end{minipage}
\begin{minipage}{.25\textwidth}
\vspace{-18em}
\includegraphics[width=\textwidth]{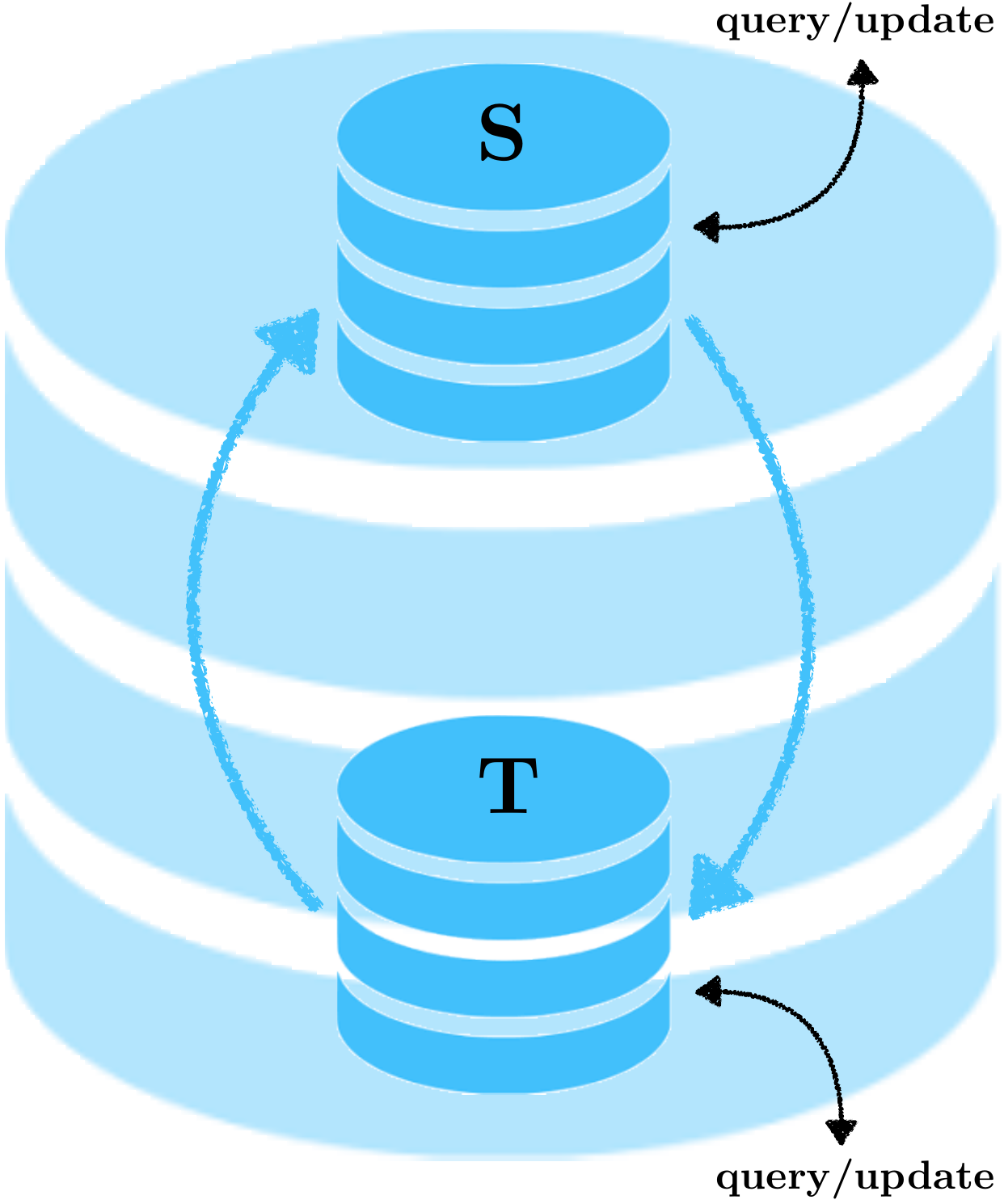}
\end{minipage}
\centering{\caption{\label{transarch}The Semantic SQL Transducer abstract architecture}}
\end{figure}

The abstract internal architecture of the Semantic SQL Transducer is shown in Figure~\ref{transarch}.
It is a generic architecture, implementing the lossless bidirectional interoperation between two databases, called \texttt{S} (source) and \texttt{T} (target). The SQL code guarantees that the two databases are always automatically synchronised after any update (wrapped within a transaction) to any of the two databases. The two databases maintain their original constraints and indexes, maximising therefore the efficiency of querying. The updates to a database are recasted directly as actual updates to the other database, maximising therefore the efficiency of updates. Some attention has to be paid to avoid infinite looping of the triggers.

The real complexity comes in defining both the lossless mappings from \texttt{S} to \texttt{T} and from \texttt{T} to \texttt{S} (appearing in the SQL code as \texttt{create table X as select ...}) and the constraints of the two databases \texttt{S} and \texttt{T}. Those mappings and constraints are provided by our theory of \emph{lossless transformations}~\cite{lubyte2009automatic,ndefo:franconi:ijsc-20,DBLP:conf/adbis/Abgrall22,DBLP:conf/aike/NdefoF19,abgrall:24}, based on the original works on \emph{information capacity}~\cite{kobayashi1986losslessness,hull1986relative,DBLP:conf/vldb/MillerIR93,DBLP:conf/edbt/Qian96,poulovassilis1998general,1260795}.

When the transducer is used to provide the semantic enrichment as a conceptual view of a data source within an ETL pipeline, the database \texttt{S} is indeed the source database, and the database \texttt{T} is the dynamic restructuring of \texttt{S} according to its conceptual schema. The involved mappings and constraints are a generalisation of classical mappings studied in the reverse engineering literature~\cite{missing:2013,lammari2007extracting,astrova2004reverse,soutou1998relational,andersson1994extracting,chiang1994reverse}, started by the seminal papers by Hainaut~\cite{hainaut2002introduction,hainaut1993contribution,hainaut93}.

We can show that any source database has associated a unique \textit{canonical abstract relational model}~\cite{DBLP:conf/ekaw/MaKOTW18}, which is the lossless materialisation of the database in its conceptual schema in 6$^{th}$ normal form. The canonical abstract relational model has a direct correspondence with the most popular conceptual modelling languages such as ERD, ORM, UML Class Diagrams, Property Graphs Schemas, RDF-based models. 
We provide a rigorous methodology to properly design the conceptual schema in the form of a canonical abstract relational model, given the data sources~\cite{DBLP:conf/adbis/Abgrall22,abgrall:24}; we are developing several tools to support it. In the next Section we will define the notion of losslessness in database restructuring, and we will explain all the basic transformation steps of the design methodology through a completely worked out example.

In conclusion, the Semantic SQL Transducer provides a seamless access to databases through their conceptual schemas,  and it contributes to a knowledge-centric data stack adding a declarative semantic layer to databases.

\section{Designing the Semantic SQL Transducer}
\label{sec:design}

In order to formalise the SQL transducer, we introduce first the notion of a \textit{first-order database schema}. A first-order database schema ${D\!B}$ is a pair $\langle\mathbb{A}_{D\!B},\mathbb{C}_{D\!B}\rangle$ where $\mathbb{A}_{D\!B}$ is a set of database predicates with their attributes $R(a_1, a_2,\cdots, a_n)$ -- for simplicity we do not consider here the domains of the attributes -- and $\mathbb{C}_{D\!B}$ is a set of first order constraints (aka dependencies) over the predicates. In order to capture exactly SQL, we restrict constraints to be in the domain-independent fragment of first-order logic: all interesting kinds of constraints can be represented, ranging from functional and multivalued dependencies (including keys), to inclusion dependencies (including foreign keys), to constraints on domain values. 
We will use a standard notation for classical database dependencies, most notably, $a_1, \cdots, a_n \rightarrow b_1,\cdots,b_m$ for functional dependencies, $a_1, \cdots, a_n \twoheadrightarrow b_1,\cdots,b_m$ for multivalued dependencies, $a_1, \cdots, a_n \subseteq b_1,\cdots,b_n$ for inclusion dependencies, and $a\subseteq\{``k_1\!",\cdots,``k_n\!"\}$ for domain constraints. We will write general dependencies using relational algebra.
%

\vspace{1ex}
\textbf{Lossless transformations.}
Suppose that $I(S)$ and $I(T)$ are the sets of legal database instances (or models) for schemas $S$ and $T$ respectively: following ~\cite{hull1986relative,DBLP:conf/vldb/MillerIR93} a \textit{schema transformation} from $S$ to $T$ is a total mapping function $f_{S \rightarrow T}: I(S) \rightarrow I(T)$. In order to define a \emph{lossless} transformation, we need to introduce first the notion of schema dominance \cite{hull1986relative,DBLP:conf/vldb/MillerIR93}.
 
Given two schemas $S$ and $T$, $T$ \textit{dominates} $S$ if there is a total and injective mapping function $f_{S \rightarrow T}: I(S) \rightarrow I(T)$ which maps legal database instances in $S$ to legal database instances in $T$. 
Equivalently, we can say that $T$ \textit{dominates} $S$ if there are two mapping functions $f_{S \rightarrow T}: I(S) \rightarrow I(T)$ and $f_{T \rightarrow S}: I(T) \rightarrow I(S)$ exist, such that their composition, $f_{T \rightarrow S}\circ f_{S \rightarrow T}$ (the result of applying $f_{T \rightarrow S}$ after applying $f_{S \rightarrow T}$), is the identity on $I(S)$.

Two schemas $S$ and $T$ are equivalent, written $S \equiv T$, if and only if $T$ dominates $S$ and $S$ dominates $T$. When two schemas $S$ and $T$ are equivalent, the mappings $f_{S \rightarrow T}$ and $f_{T \rightarrow S}$ are bijective, and we say that both schemas have the same \emph{information capacity} and that the transformation is \emph{lossless}.

In our setting, we consider mappings as first-order \emph{views} establishing the relation between two database schemas. More precisely, given two schemas $S$ and $T$, a first-order mapping from $S$ to $T$, written $M_{S\rightarrow T}$, is a set of first-order views $R_T = e_S^{R_T}$ for each predicate $R_T$ of arity $n$ in $\mathbb{A}_T$, with $e_S^{R_T}$ a relational algebra expression of arity $n$ over the alphabet $\mathbb{A}_S$.

In the first-order setting, it can be proved that the definition of equivalence (or lossless transformation) between $S$ and $T$ corresponds to the following condition over the schemas and mappings: $(\mathbb{C}_S \cup M_{S\rightarrow T}) \equiv (\mathbb{C}_T \cup M_{T\rightarrow S}) $, which really means $I(\langle\mathbb{A}_S\cup\mathbb{A}_T,\mathbb{C}_S\cup M_{S\rightarrow T}\rangle) = I(\langle\mathbb{A}_S\cup\mathbb{A}_T,\mathbb{C}_T\cup M_{T\rightarrow S}\rangle)$; see~\cite{DBLP:conf/aike/NdefoF19,ndefo:franconi:ijsc-20}.

\vspace{1ex}
\textbf{Transformation patterns.}
\emph{Transformation patterns} are crafted templates describing a specific structure of schema transformation with the constraints necessary to ensure its losslessness~\cite{abgrall:24}.
The two basic lossless transformation patterns are \emph{vertical decomposition} and \emph{horizontal decomposition}.
We introduce them via two basic examples.

Given two schemas $S$ and $T$ as follows:

\vspace{1ex}
$S=\langle\{p(a,b,c)\},(p.b\rightarrow p.c)\}\rangle$

$T=\langle\{q(a,b),r(b,c)\},(r.b\rightarrow r.c), (q.b = r.b)\}\rangle$

\vspace{1ex}
\noindent
The schemas $S$ and $T$ have the same information capacity since there is a lossless transformation through the following mappings -- characterising the vertical decomposition in the classical database literature:

\vspace{1ex}
$M_{S\rightarrow T}=\{(q=\pi_{ab}\ p), (r=\pi_{bc}\ p)\}$

$M_{T\rightarrow S}=\{(p=q\bowtie r)\}$

\vspace{1ex}
\noindent
The vertical decomposition transformation pattern maps a schema with a join dependency (i.e., a key dependency, or a functional dependency, or a multivalued dependency) to its vertical decomposition, with all the appropriate dependencies in both schemas to guarantee losslessness.

As an example of a lossless horizontal decomposition transformation, consider the schema $U$:

\vspace{1ex}
$U=\langle\{q(a,b),r_1(b,c),r_2(b,c)\},
\{$%
\parbox[t]{\textwidth}{$%
(r_1.b\rightarrow r_1.c),(r_2.b\rightarrow r_2.c), \\
(r_1.c = \{``k"\}), (r_2.c \not\subseteq \{``k"\}),\\
(\pi_b\ q = \pi_b\ r_1 \cup {}\pi_b\ r_2),(\pi_b\ r_1 \cap {}\pi_b\ r_2=\emptyset)
\}\rangle$}

\vspace{1ex}
\noindent
The schemas $T$ and $U$ have the same information capacity since there is a lossless transformation through the following mappings -- characterising the horizontal decomposition via the condition $\sigma_{c =``k"}\ r$\ :

\vspace{1ex}
$M_{T\rightarrow U}=\{(r_1=\sigma_{c =``k"}\ r), (r_2=\sigma_{c \neq``k"}\ r)\}$

$M_{U\rightarrow T}=\{(r=r_1\cup r_2)\}$

\vspace{1ex}
\noindent
The horizontal decomposition transformation pattern maps a schema to a horizontally decomposed one via a selection condition, with all the appropriate dependencies in both schemas to guarantee losslessness. 

We can also observe that also $S$ and $U$ have the same information capacity, since they are related by a sequence of lossless transformations.

A special case of horizontal decomposition is the lossless transformation leading to a SQL NULL-free schema. According to the logic theory of SQL NULL values~\cite{DBLP:journals/corr/abs-2202-10898,DBLP:conf/amw/FranconiT12}, a schema has the same information capacity as an horizontally decomposed one via a NULLABLE condition over some attribute. Whenever there is a NULLABLE constraint over an attribute, a table can be losslessy decomposed into two tables, one having all the attributes but not the NULLABLE one, and the other having all the attributes but with a NOT NULL constraint replacing the NULLABLE constraint.

We have identified several lossless transformation patterns~\cite{ndefo:franconi:ijsc-20,abgrall:24}, which can be used to design a Semantic SQL Transducer allowing for arbitrary data restructuring processes, whenever we want to guarantee that no information is lost during the restructuring process. The transformation patterns identify the lossless mappings from \texttt{S} to \texttt{T} and from \texttt{T} to \texttt{S} and the constraints of the two databases \texttt{S} and \texttt{T}, needed to design a correctly working Semantic SQL Transducer, as described in Section~\ref{sec:inside}.

A very special data restructuring process is the \emph{reverse engineering} process, which looks for the lossless transformation from a source database schema to the schema corresponding to its conceptual schema -- see~\cite{hainaut2002introduction} for a survey. This is the scenario we have presented in Section~\ref{sec:role}: we want to expose the source data with a vocabulary that corresponds to its conceptual schema, useful for the business perspective. If the transformation is lossless, we have the guarantee that no information is lost, and that high-level users can query and update freely the transformed database, in this case the database organised in a meaningful structure. More specifically, we want to losslessly transform a source schema into a schema in 6$^{th}$ normal form with explicit \emph{Object Identifiers} (OIDs) to identify ``entities", namely instances of entity types. Object identifiers can be implemented by surrogate keys, URIs, or UUIDs. This form is called an Abstract Relational Model (ARM)~\cite{DBLP:conf/er/BorgidaTW16}. We can show that (a) given an arbitrary database schema, there exists a unique \emph{Canonical Abstract Relational Model} (CARM) for that schema, based on the 5$^{th}$ or 6$^{th}$ normal forms, which plays the role of the \emph{Core Conceptual Schema} of the original database, and (b) the CARM schema corresponds to a unique conceptual schema directly expressible in conceptual modelling languages such as ORM, EER, UML class diagrams, or in RDF-based modelling languages.


In order to understand how to losslessly transform a database schema into an equivalent one (the CARM) which includes OIDs, let's consider the following basic example.
Assume we have a schema in 5$^{th}$ normal form, so that the only constraints within a table are key constraints, and the constraints across tables are foreign keys or inverse foreign keys, for example:
 
\vspace{1em}
\textsf{Employee(ssn,name), works-in(ssn,depname), Department(depname,address)}
 
\textsf{Employee.ssn $\rightarrow$ Employee.name}

\textsf{works-in.ssn $\rightarrow$ Employee.ssn}

\textsf{works-in.depname $\rightarrow$ Department.depname}

\textsf{Department.depname $\rightarrow$ Department.address}

\vspace{1em}
\noindent
A domain expert should recognise that \textsf{Employee} and \textsf{Department} are \emph{entity types}, while \textsf{works-in} is a \textit{relationship type}. As a rule of thumb, we can recognise entity types since they should be the target of at least a foreign key with a relationship type as source, while a relationship type should have at least one attribute as the source of a foreign key with an entity type as target. A new attribute with domain OID (disjoint from STRING and INTEGER) is added as a surrogate key to each entity type table, and coherently a new OID attribute replaces the attributes involved in a foreign key path from the entity type. The foreign key and inverse foreign key constraints holding across tables are duplicated to hold between the added OID attributes. Following our example, the lossless transformation of the above schema with added OIDs is:

\vspace{1em}
\textsf{Employee(eoid,ssn,name), works-in(eoid,doid), Department(doid,depname,address)}

\textsf{Employee.eoid $\rightleftarrows$ Employee.ssn}

\textsf{Employee.ssn $\rightarrow$ Employee.name}

\textsf{works-in.eoid $\rightarrow$ Employee.eoid}

\textsf{works-in.doid $\rightarrow$ Department.doid}

\textsf{Department.doid $\rightleftarrows$ Department.depname}

\textsf{Department.depname $\rightarrow$ Department.address}

\vspace{1em}
\noindent
The schema above is the logical representation of the diagrammatic conceptual schema below, expressed here as an Entity-Relationship Diagram:

\vspace{1ex}
\begin{center}
	\includegraphics[width=.75\textwidth]{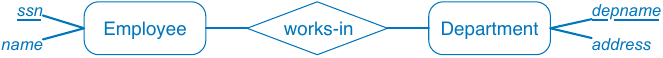}
\end{center}

\begin{figure}[t]
\includegraphics[width=\textwidth]{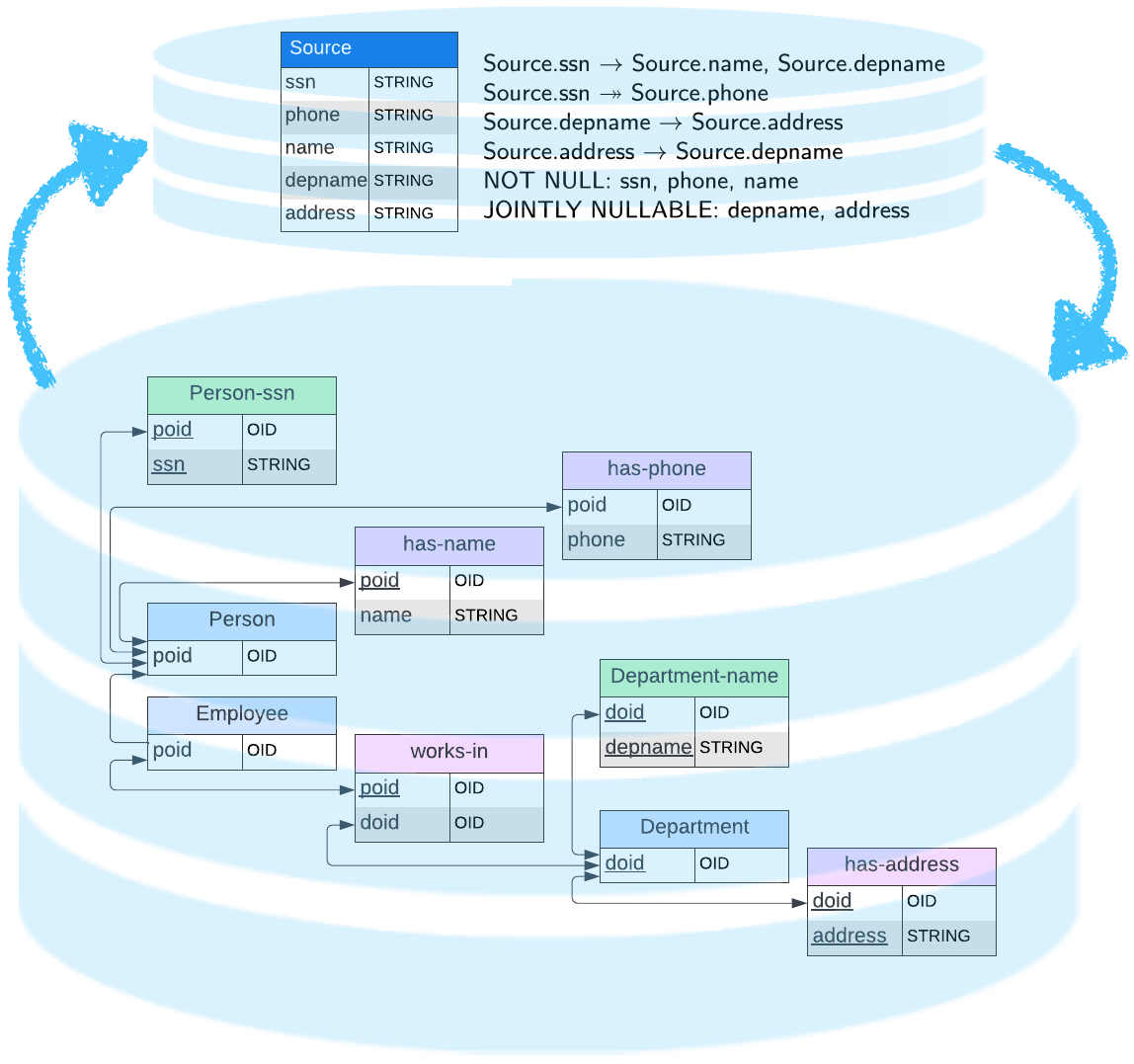}
\centering{\caption{\label{fig:ex-transf}The Semantic SQL Transducer of the example}}
\end{figure}

\vspace{1ex}
\textbf{A complete example.}
Suppose we have a source database schema as described at the top of Figure~\ref{fig:ex-transf}. The schema is composed by a single table \textsf{Source} and a set of constraints. As humans, we can tell that the schema is about people, identified by their social security number, having one or more phone numbers, and possibly working in some department, identified by its name, having an address, which also uniquely identifies the department at that address.

\begin{figure}[t]
\includegraphics[width=.8\textwidth]{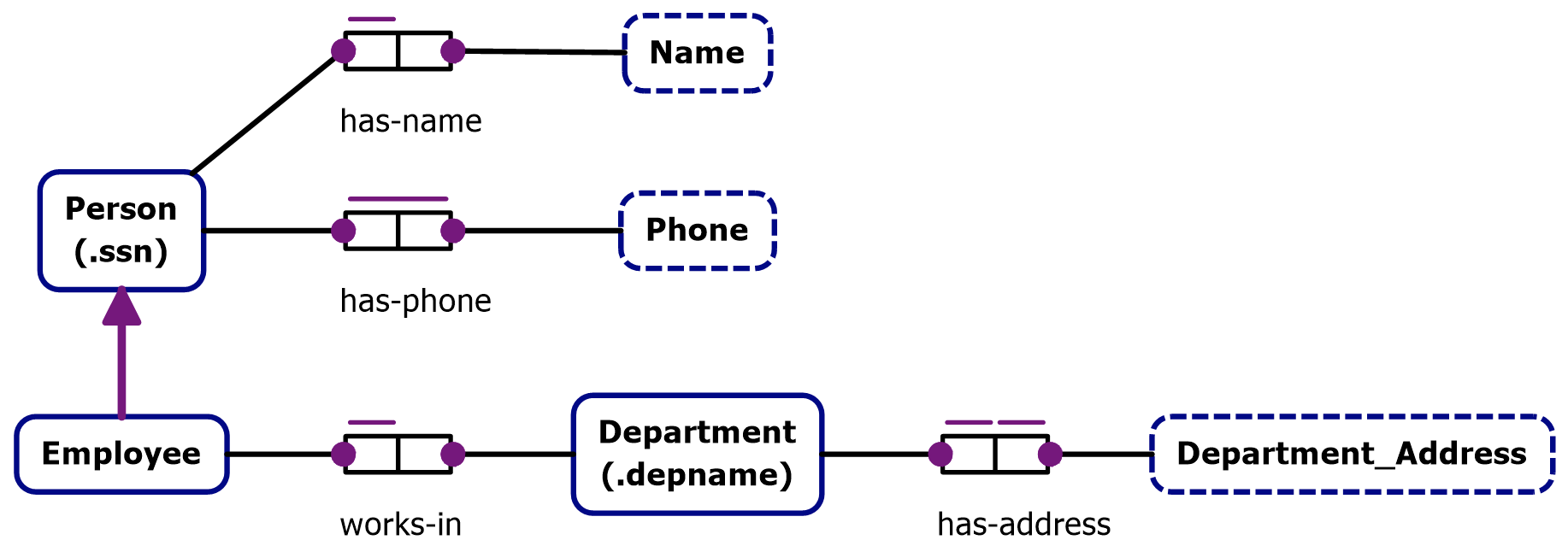}
\centering{\caption{\label{fig:conc}The conceptual schema of the example (in the ORM notation)}}
\end{figure}

Clearly, a lot of the information we just described about this database is hidden in the schema, and we may wonder which could be its explicit conceptual schema to have a more direct understanding of the data. By jumping a little ahead, let's have a glance at the conceptual schema, expressed in the ORM notation, in Figure~\ref{fig:conc}. That schema describes precisely, non-ambiguously, and formally the data as we were saying above. But how can we get this conceptual schema from the original source schema? First we notice that this conceptual schema has a direct representation as a logical schema in the relational setting: this is described at the bottom of Figure~\ref{fig:ex-transf}. The relational version of the conceptual schema denotes exactly the same legal databases as the ORM schema of Figure~\ref{fig:conc}, but now in a pure relational setting. Note that the constraints of this schema are only key constraints and unary inclusion dependencies among OID datatypes -- this property holds for all core conceptual schemas derivable from arbitrary source schemas.

The core conceptual schema is obtained by the application of a sequence of lossless transformation patterns, of the type we have briefly introduced in the previous Section. Since any lossless transformation step is accompanied by the mappings in the two directions, we get from the transformation process also the views from the source data to the conceptual data \emph{and} the views from the conceptual data to the source data. In our example, these mappings are shown in Figure~\ref{fig:mappings}. We have devised a methodology driving the design of the correct sequence of lossless transformation patterns leading to a core conceptual schema from a source schema.

\begin{figure}[b]
{\small
$
\begin{array}{rl}
\textsf{Person}      &=\varrho_\textsf{\scriptsize [ssn/poid]}(\pi_\textsf{\scriptsize ssn}(\textsf{Source}))\\
\textsf{Person-ssn}  &=\varrho_\textsf{\scriptsize [ssn$_1$/poid,ssn$_2$/ssn]}(\sigma_\textsf{\scriptsize ssn$_1$=ssn$_2$}(\pi_\textsf{\scriptsize ssn$_1$,ssn$_2$}(\textsf{Source}\times\textsf{Source})))\\
\textsf{has-name} &=\varrho_\textsf{\scriptsize [ssn/poid]}(\pi_\textsf{\scriptsize ssn,name}(\textsf{Source}))\\
\textsf{has-phone}&=\varrho_\textsf{\scriptsize [ssn/poid]}(\pi_\textsf{\scriptsize ssn,phone}(\textsf{Source}))\\
\textsf{Employee} &=\varrho_\textsf{\scriptsize [ssn/poid]}(\pi_\textsf{\scriptsize ssn}(\sigma_{\textsf{\scriptsize depname NOT NULL}}\textsf{Source}))\\
\textsf{works-in} &=\varrho_\textsf{\scriptsize [ssn/poid]}(\pi_\textsf{\scriptsize ssn,depname}(\sigma_{\textsf{\scriptsize depname NOT NULL}}\textsf{Source}))\\
\textsf{Department}      &=\varrho_\textsf{\scriptsize [depname/doid]}(\pi_\textsf{\scriptsize depname}(\sigma_{\textsf{\scriptsize depname NOT NULL}}\textsf{Source}))\\
\textsf{Department-name} &=\varrho%
\begin{minipage}[t]{\textwidth}$%
_\textsf{\scriptsize [depname$_1$/doid,depname$_2$/depname]}\\
(\sigma_\textsf{\scriptsize depname$_1$=depname$_2$}(\pi_\textsf{\scriptsize depname$_1$,depname$_2$}$\\
\begin{minipage}[t]{\textwidth}
$%
~~~((\sigma_{\textsf{\scriptsize depname NOT NULL}}\textsf{Source})\times(\sigma_{\textsf{\scriptsize depname NOT NULL}}\textsf{Source}))))
$\end{minipage}
\end{minipage}\\
\textsf{has-address}&=\varrho_\textsf{\scriptsize [depname/doid]}(\pi_\textsf{\scriptsize depname,address}(\sigma_{\textsf{\scriptsize depname NOT NULL}}\textsf{Source}))\\
\\
\textsf{Source}         &=
\pi
\begin{minipage}[t]{\textwidth}
$_\textsf{\scriptsize ssn,phone,name,depname,address}$\\
$\textbf{\large $\fullouterjoin$}~ $
(\textsf{Person, Employee, has-phone, has-name,} \\
\begin{minipage}[t]{\textwidth}
~~~~~~~~~\textsf{works-in, has-address, Person-ssn, Department-name})
\end{minipage}
\end{minipage}
\end{array}
$
}
\centering{\caption{\label{fig:mappings}The lossless mappings from source to CARM and viceversa}}
\end{figure}

So now we have all the ingredients (the two sets of constraints and the two mappings) to finalise the implementation of a Semantic SQL Transducer for the source schema as explained in Section~\ref{sec:design}. Many things can be done once we have the transducer in place. We can understand what the source data is about. We can query the source database with SQL queries using only the core conceptual schema table names, bridging therefore the gap between business models and available data sources. Along these lines, business users can directly update the source data using just the conceptual vocabulary. In the other direction, legacy queries over the source database can be explained by looking at their expansion in terms of the conceptual schema. The Semantic SQL Transducer provides a well-founded semantic layer for data catalogues. Source data structured according its core conceptual schema makes data analysis much more effective, since the correlations, the classifications, and the similarities of the data elements are much more meaningful when done in business terms.

\section{Conclusions}
We have introduced in this paper a tool to losslessly restructure relational data, allowing for seamless views of the data, which can be queried and updated at both ends maintaining consistency and integrity. A special kind of transformation is when a database is restructured according to its conceptual schema, providing therefore a materialised copy of the data, always in sync with it, using the actual vocabulary understood by the business.

\begin{credits}
\subsection*{\ackname}
This long-standing work has been realised through collaborations and discussions with Nicola Pedot, Nonyelum Ndefo, Francesco Sportelli, Sergio Tessaris, Volha Kerhet, Nhung Ngo, Paolo Guagliardo, David Toman, Grant Weddell, Alex Borgida, Terry Halpin, Jan Hidders, Sebastian Link. The activity has been partly funded by the Confucius project of the Free University of Bozen-Bolzano.
\end{credits}
\vfill
\pagebreak

\nocite{*}
\bibliographystyle{splncs04}
\bibliography{caise-ws-24.bib}

\begin{thebibliography}{10}
\providecommand{\url}[1]{\texttt{#1}}
\providecommand{\urlprefix}{URL }
\providecommand{\doi}[1]{https://doi.org/#1}

\bibitem{DBLP:conf/adbis/Abgrall22}
Abgrall, T.: Formalization of data integration transformations. In: New Trends
  in Database and Information Systems - {ADBIS} 2022. vol.~1652, pp. 615--622.
  Springer (2022)

\bibitem{abgrall:24}
Abgrall, T.: Schema decomposition via transformation patterns (2024), submitted

\bibitem{abgrall:caise-forum-24}
Abgrall, T., Franconi, E., Pedot, N.: Enhancing data management and value
  creation through a knowledge-centric data stack (2024), submitted

\bibitem{DBLP:journals/jcss/AlbertIR99}
Albert, J., Ioannidis, Y.E., Ramakrishnan, R.: Equivalence of keyed relational
  schemas by conjunctive queries. J. Comput. Syst. Sci.  \textbf{58}(3),
  512--534 (1999)

\bibitem{andersson1994extracting}
Andersson, M.: Extracting an entity relationship schema from a relational
  database through reverse engineering. In: International Conference on
  Conceptual Modeling. pp. 403--419. Springer (1994)

\bibitem{astrova2004reverse}
Astrova, I.: Reverse engineering of relational databases to ontologies. In:
  European Semantic Web Symposium. pp. 327--341. Springer (2004)

\bibitem{DBLP:conf/er/BorgidaTW16}
Borgida, A., Toman, D., Weddell, G.E.: On referring expressions in information
  systems derived from conceptual modelling. In: Conceptual Modeling - 35th
  International Conference, {ER} 2016. pp. 183--197 (2016)

\bibitem{DBLP:books/sp/18/CalvaneseF18}
Calvanese, D., Franconi, E.: First-order ontology mediated database querying
  via query reformulation. In: A Comprehensive Guide Through the Italian
  Database Research Over the Last 25 Years, vol.~31, pp. 169--185. Springer
  (2018)

\bibitem{DBLP:reference/db/CalvaneseGLLR18}
Calvanese, D., Giacomo, G.D., Lembo, D., Lenzerini, M., Rosati, R.:
  Ontology-based data access and integration. In: Liu, L., {\"{O}}zsu, M.T.
  (eds.) Encyclopedia of Database Systems, Second Edition. Springer (2018)

\bibitem{chiang1994reverse}
Chiang, R.H., Barron, T.M., Storey, V.C.: Reverse engineering of relational
  databases: Extraction of an {EER} model from a relational database. Data \&
  knowledge engineering  \textbf{12}(2),  107--142 (1994)

\bibitem{DBLP:conf/itadata/CiacciaMT22}
Ciaccia, P., Martinenghi, D., Torlone, R.: Conceptual constraints for data
  quality in data lakes. In: Proceedings of the 1st Italian Conference on Big
  Data and Data Science (itaDATA 2022), Milan, Italy, September 20-21, 2022.
  pp. 111--122 (2022)

\bibitem{DBLP:journals/jdiq/ConsoleL23}
Console, M., Lenzerini, M.: Editorial: Special issue on quality aspects of data
  preparation. {ACM} J. Data Inf. Qual.  \textbf{15}(4),  40:1--40:2 (2023)

\bibitem{DBLP:journals/semweb/Cudre-Mauroux20}
Cudr{\'{e}}{-}Mauroux, P.: Leveraging knowledge graphs for big data
  integration: the {XI} pipeline. Semantic Web  \textbf{11}(1),  13--17 (2020)

\bibitem{DBLP:conf/semweb/Dibowski0SHT20}
Dibowski, H., Schmid, S., Svetashova, Y., Henson, C., Tran, T.: Using semantic
  technologies to manage a data lake: Data catalog, provenance and access
  control. In: 12th International Workshop on Scalable Semantic Web Knowledge
  Base Systems. pp. 65--80 (2020)

\bibitem{DBLP:conf/dmdw/FranconiS99}
Franconi, E., Sattler, U.: A data warehouse conceptual data model for
  multidimensional aggregation. In: Intl. Workshop on Design and Management of
  Data Warehouses, DMDW'99 (1999)

\bibitem{DBLP:conf/amw/FranconiT12}
Franconi, E., Tessaris, S.: On the logic of {SQL} nulls. In: 6th Alberto
  Mendelzon International Workshop on Foundations of Data Management. pp.
  114--128. CEUR Workshop Proceedings (2012)

\bibitem{DBLP:journals/corr/abs-2202-10898}
Franconi, E., Tessaris, S.: Relational algebra and calculus with {SQL} null
  values. CoRR  \textbf{abs/2202.10898} (2022)

\bibitem{hainaut1993contribution}
Hainaut, J.L., Chandelon, M., Tonneau, C., Joris, M.: Contribution to a theory
  of database reverse engineering. In: [1993] Proceedings Working Conference on
  Reverse Engineering. pp. 161--170. IEEE (1993)

\bibitem{hainaut2002introduction}
Hainaut, J.L.: Introduction to database reverse engineering. LIBD Lecture Notes
   (2002)

\bibitem{hainaut93}
Hainaut, J.L., Tonneau, C., Joris, M., Chandelon, M.: Schema Transformation
  Techniques for Database Reverse Engineering, pp. 353--372. Springer Verlag
  (1993)

\bibitem{DBLP:journals/sigmod/HameedN20}
Hameed, M., Naumann, F.: Data preparation: {A} survey of commercial tools.
  {SIGMOD} Rec.  \textbf{49}(3),  18--29 (2020)

\bibitem{hull1986relative}
Hull, R.: Relative information capacity of simple relational database schemata.
  SIAM Journal on Computing  \textbf{15}(3),  856--886 (1986)

\bibitem{10.14778/3574245.3574274}
Khatiwada, A., Shraga, R., Gatterbauer, W., Miller, R.J.: Integrating data lake
  tables. Proc. VLDB Endow.  \textbf{16}(4),  932--945 (dec 2022)

\bibitem{kobayashi1986losslessness}
Kobayashi, I.: Losslessness and semantic correctness of database schema
  transformation: another look of schema equivalence. Information Systems
  \textbf{11}(1),  41--59 (1986)

\bibitem{lammari2007extracting}
Lammari, N., Comyn-Wattiau, I., Akoka, J.: Extracting generalization
  hierarchies from relational databases: A reverse engineering approach. Data
  \& Knowledge Engineering  \textbf{63}(2),  568--589 (2007)

\bibitem{lubyte2009automatic}
Lubyte, L., Tessaris, S.: Automatic extraction of ontologies wrapping
  relational data sources. In: International Conference on Database and Expert
  Systems Applications. pp. 128--142. Springer (2009)

\bibitem{DBLP:conf/ekaw/MaKOTW18}
Ma, W., Keet, C.M., Oldford, W., Toman, D., Weddell, G.E.: The utility of the
  abstract relational model and attribute paths in {SQL}. In: Knowledge
  Engineering and Knowledge Management - 21st International Conference, {EKAW}.
  pp. 195--211. Springer (2018)

\bibitem{1260795}
McBrien, P., Poulovassilis, A.: Data integration by bi-directional schema
  transformation rules. In: Proceedings 19th International Conference on Data
  Engineering (ICDE-03). pp. 227--238 (2003)

\bibitem{missing:2013}
Mian, N.A., Khan, S.A., Zafar, N.A.: Database reverse engineering methods: What
  is missing? Res. J. Recent Sci.  \textbf{2}(5),  49--58 (May 2013)

\bibitem{DBLP:conf/vldb/MillerIR93}
Miller, R.J., Ioannidis, Y.E., Ramakrishnan, R.: The use of information
  capacity in schema integration and translation. In: 19th International
  Conference on Very Large Data Bases (VLDB-1993). pp. 120--133 (1993)

\bibitem{DBLP:conf/aike/NdefoF19}
Ndefo, N., Franconi, E.: On preserving information in schema transformations:
  {A} constructive perspective. In: 2nd {IEEE} International Conference on
  Artificial Intelligence and Knowledge Engineering, {AIKE} 2019. pp. 57--64
  (2019)

\bibitem{ndefo:franconi:ijsc-20}
Ndefo, N., Franconi, E.: A study on information-preserving schema
  transformations. International Journal of Semantic Computing  \textbf{14}(1),
   27--53 (2020)

\bibitem{PAPOTTI2009665}
Papotti, P., Torlone, R.: Schema exchange: Generic mappings for transforming
  data and metadata. Data \& Knowledge Engineering  \textbf{68}(7),  665--682
  (2009)

\bibitem{poulovassilis1998general}
Poulovassilis, A., McBrien, P.: A general formal framework for schema
  transformation. Data \& Knowledge Engineering  \textbf{28}(1),  47--71 (1998)

\bibitem{DBLP:conf/edbt/Qian96}
Qian, X.: Correct schema transformations. In: EDBT'96, 5th International
  Conference on Extending Database Technology. pp. 114--128. Springer (1996)

\bibitem{soutou1998relational}
Soutou, C.: Relational database reverse engineering: algorithms to extract
  cardinality constraints. Data \& Knowledge Engineering  \textbf{28}(2),
  161--207 (1998)

\end{thebibliography}

\end{document}